\documentclass[runningheads]{llncs}
\usepackage{graphicx}
\begin{document}
\title{Leveraging Semantic Relationships to Prioritise Indicators of Compromise in Additive Manufacturing Systems}
%
%
\author{Mahender Kumar\inst{1} \and
Gregory Epiphaniou\inst{2} \and
Carsten Maple\inst{3}}
\authorrunning{Kumar et al.}
%
\institute{Cyber Security Research Group, WMG, University of Warwick \\
Coventry, United Kingdom \\
\email{Mahender.kumar@warwick.ac.uk \and Gregory.epiphaniou@warwick.ac.uk \and CM@warwick.ac.uk}
}
\maketitle              
\begin{abstract}
Additive manufacturing (AM) offers numerous benefits, such as manufacturing complex and customised designs quickly and cost-effectively, reducing material waste, and enabling on-demand production. However, several security challenges are associated with AM, making it increasingly attractive to attackers ranging from individual hackers to organised criminal gangs and nation-state actors. This paper addresses the cyber risk in AM to attackers by proposing a novel semantic-based threat prioritisation system for identifying, extracting and ranking indicators of compromise (IOC). The system leverages the heterogeneous information networks (HINs) that automatically extract high-level IOCs from multi-source threat text and identifies semantic relations among the IOCs. It models IOCs with a HIN comprising different meta-paths and meta-graphs to depict semantic relations among diverse IOCs. We introduce a domain-specific recogniser that identifies IOCs in three domains: \textit{organisation\_specific}, \textit{regional\_source-specific}, and \textit{regional\_target-specific}. A threat assessment uses similarity measures based on meta-paths and meta-graphs to assess semantic relations among IOCs. It prioritises IOCs by measuring their severity based on the frequency of attacks, IOC lifetime, and exploited vulnerabilities in each domain. 

\end{abstract}
\section{Introduction }

Industry 4.0, the fourth industrial revolution, refers to integrating advanced digital technologies and manufacturing systems to automate and optimize industrial processes. Additive manufacturing (AM) is a key enabler of Industry 4.0, as it allows for the rapid and flexible production of customized parts and products [1]. AM is a process that enables the production of complex devices by applying successive layers of materials. AM offers many advantages, such as on-demand customisation, enhanced logistics, reduced labour and production lead times, streamlined production, reduced waste, reduced inventory, and reduced transportation costs. However, cyber and physical attacks in AM pose severe concerns and formidable challenges [2], making AM supply chains susceptible to various attack vectors. As a result, protecting the security of AM has become increasingly important, and developing robust security mechanisms that protect against a range of potential attacks has become a significant challenge for researchers and industry practitioners alike. 

Modern attacks on AM are often sophisticated and can exploit hidden vulnerabilities that go undetected for long periods [3-7]. A prime example is Advanced Persistent Threats (APTs), which have been used to target AM industries for espionage, economic gain, and intellectual property theft. APTs are commonly described as an extended attack campaign in which one or more intruders execute a long-term plan to take control of a network or system. In 2020, more than 1000 data breaches were reported in the United States alone, affecting more than 155.8 million individuals through data exposure [3]. Perhaps the most famous kinetic cyber attack of all time was aimed at Iran's nuclear program, considered unprecedented in the industry [4]. The Stuxnet attack involved a complexly targeted worm that executed zero-day exploits on operating systems and software for managing programmable logic controllers (PLCs). The attack resulted in tens of billions of dollars in damage. Another famous example of a cyberattack is the sewage attack in Maroochy Shire, which caused a system failure and millions of litres of untreated sewage to leak into the water supply [5]. Belikovetsky et al. [6] conducted a study on the vulnerability of additive manufacturing to cyber attacks. He demonstrated the sabotage attack on a propeller blueprint that can be 3D printed at home. The findings of their study emphasized the vulnerability of additive constructs to cyber attacks, which was also confirmed by another recent paper [7]. The authors of the latter paper identified AM as the second most vulnerable industry to cyberattacks, second only to the financial sector. With AM's growing national and industrial importance, cyberattacks have become more attractive and induced threat actors are increasingly involved in cybercrime-as-a-service, commoditising cyberattacks. As a result, APTs are now employing common attack patterns to compromise targets. Therefore, early identification of threat exposure and breaches is critical to preventing significant damage to an organization and providing reliable evidence during prosecution trials.

Cyber Threat Intelligence (CTI) can be a valuable tool for assessing the threat landscape in AM and developing effective strategies for mitigating cyber risks. Threat intelligence feeds can help organizations stay informed about emerging threats and new attack techniques. This can be especially useful in the fast-paced world of AM, where new technologies and processes are constantly being developed. CTI involves the extraction of threat intelligence from threat-related information from multiple sources, utilising several attributes, including Indicators of Compromise (IOCs), Tactic, Technique, and Procedure (TTP) and the skills and motive of the threat actor [8]. Some example of CTI feeds are Structured Threat Information Expression (STIX), OpenDef, Cybox, and OpenIOC, but a massive amount of information remain unstructured. IBM X-force [9], Facebook ThreatExchange [10], OpenCTI [11] and MISP [12] are a few vendors who provide threat intelligence feeds by extracting threat intelligence from multiple open sources using IOC extraction methods such as PhishTank, and IOCFinder.

These structured threat information feeds have several disadvantages, including a limited scope, delayed information, high cost, inflexibility and false positives, making it very challenging for AM industry to rely on them. On the other hand, unstructured threat feeds can provide a more comprehensive and flexible approach to threat intelligence that can be more effective for many organizations. However, unstructured reports may not be well-organized, making it hard to identify the relationships between different pieces of information. Other challenges include errors, inaccuracies, and missing information. As a result, it requires advanced natural language processing (NLP) techniques and machine learning algorithms to extract meaningful and relevant threat information from unstructured reports.

\begin{figure}
\includegraphics[width=\textwidth]{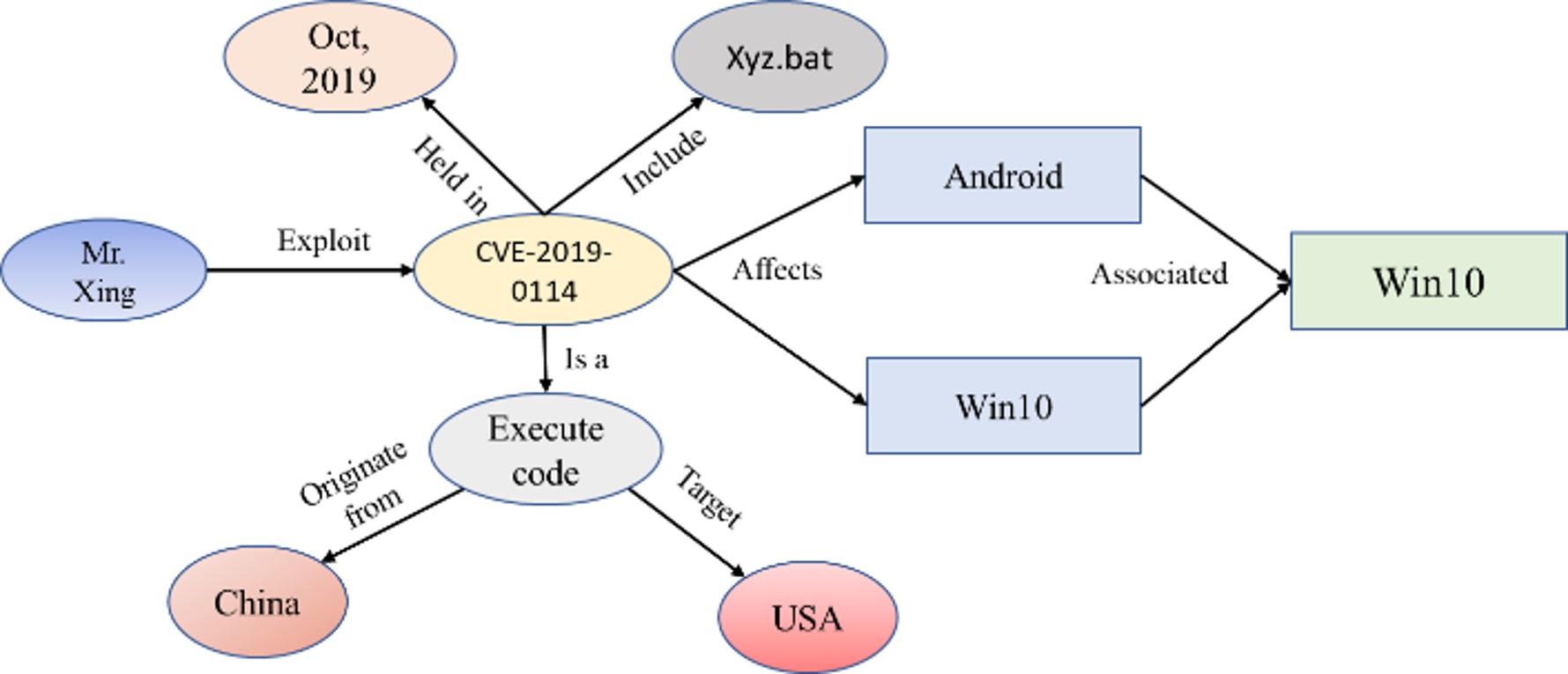}
\caption{An annotated example of CTI includes IOCs such as attack actor, vulnerability, time, region, file, attack type, device and platform, and their relationship.} \label{fig1}
\end{figure}

 Consider the following instance of a security-related post: "\textit{In October 2019, Mr Xing from China exploited the CVE-2019-0114 vulnerability, which affected multiple Android and Win10 devices in the United States. CVE-2019-0114 is a remote code execution vulnerability that contains the malicious file abc.bat}". Figure 1 displays a graphical representation of CTI, including eight IOCs such as attack actor, vulnerability, time, region, file, attack type, device and platform, and the relationship between them. Existing methods only consider IOCs but avoid the relationship between them, and as a result, they cannot grasp a comprehensive picture of the threat landscape. To overcome the limitations of existing structured threat feed tools, this paper aims to automate IOC extraction by exploiting Heterogeneous Information Network (HIN) that provides insight into the interdependencies between heterogeneous IOCs.

This paper presents a novel semantic-based threat prioritisation framework for identifying, extracting and ranking IOCs. The summary of the paper is as follows:

\begin{itemize}
    \item \textit{Recogniser}. We propose a recogniser that automatically extracts threat-related information from multi-source threat text. It also identifies the domains to which IOCs belong and integrates IOCs with their domain, forming three domain-specific IOCs such as \textit{organisation\_domain-specific}, \textit{regional\_source-specific}, and \textit{regional\_target-specific} threat intelligence. 
    \item \textit{Threat modelling}. We model the range of IOCs with a Heterogeneous Information Network (HIN), which comprises different meta-paths and meta-graphs that depicts the semantic relations among diverse IOCs to capture a more comprehensive landscape of threat events. 
    \item \textit{Threat assessment}. We present a CTI assessment framework that uses similarity measures based on meta-paths and meta-graphs and assesses the interdependent relations among diverse IOCs. 
    \item \textit{Prioritisation}. We then measure the severity of IOCs by considering the frequency of attacks, IOC lifetime, and the number of exploited vulnerabilities in each domain. As a result, they evaluate the ranking mechanism for each IOC.
\end{itemize}

The rest of the paper is organized as follows: Section 2 discusses the related work, and Section 3 provides the conceptual background. The proposed framework is presented in Section 4. Finally, Section 5 summarizes the paper and provides directions for future research.

\section{Related Work}

Extracting threat intelligence from the unstructured text of threat-related information has become an exciting research topic in cyber security. This section briefly describes key methodologies for identifying cyber threats by extracting IOCs from multiple sources.

Noor et al. [13] have proposed a model to automate cyber threat attribution by considering high-level IOCs to determine threat actors. Their technique extracts high-level IOCs from unstructured CTI reports and then semantically profiles threat actors with the high-level IOCs taken from MITRE’s ATT\&CK. Zhao et al. [14] present TIMiner, a method to extract and assess domain-specific CTIs that automatically classify the domains associated with related CTIs. Gao et al. [15] proposed a cyber threat intelligence method based on the Hierarchical Information Network (HINCTI) system to identify threat types. HINCTI provides a threat intelligence-based meta-schema to capture the semantic relationship between threat nodes leveraging a meta-path and meta-graph-based similarity method. Zhao et al. [16] proposed a CTI framework, HINTI, based on HIN that proposed a multi-granular-based IOC recogniser to enhance the accuracy of the IOC extraction method. HINTI defines different types of IOCs using meta-paths to identify the relationship between IOCs, profile the threat events and rank the significance of IOCs to understand the threat landscape.

Liao et al. [17] proposed a novel automated threat-related information collection and extraction (iACE) technique from unstructured text that uses a natural language processing method to extract IOC data from text documents and then analyse the IOC data using graph mining methods. iACE aims to identify the grammatical semantic relationships between token threat patterns associated with IOC in text documents. The method integrates name entity recognition and relation extraction methods. Gao et al. [18] proposed a threat-hunting system (THREATRAPTOR) that extracts threat behavioural information from unstructured CTI reports to facilitate threat hunting. THREATRAPTOR provides an accurate NLP-based auditing framework to extract structured threat information from unstructured CTI text and defines a domain-specific query language to detect malicious system activities. Wang et al. [19] develop an efficient automated process that recognises and extracts entities and their relationship from text reports.

\section{Conceptual Background}

\subsection{Cyber Threat Intelligence}

Modern cybercriminals have developed sophisticated tactics, techniques, and procedures (TTP) to realise their aim of compromising their targets quickly and efficiently. Thus, traditional defence mechanisms, such as anti-virus software, firewalls and intrusion detection methods, struggle to effectively detect cyber attacks such as advance persistent threats (APTs) and zero-day attacks. Cyber attacks have successfully compromised systems in a wide range of sectors. For example, the WannaCry ransomware attack extorted money to unlock sensitive information and designs across various industries [20]. Security experts have increasingly turned to sharing cyber threat intelligence (CTI) to combat such emerging cyber threats. CTI is any relevant information that helps detect, monitor, assess, and respond to cyber threats. CTI facilitates a comprehensive and significant threat warning and includes information such as IOCs [21]. 

Nowadays, a rich source of commercial and free CTI feeds are available, making it difficult for network defenders to evaluate the quality of information and select the optimal set of data feeds to pay attention to. Acting on results from low-quality feeds can give rise to many false alerts while concentrating on only a few data feeds increases the risk of missing relevant threats. However, it is challenging to extract IOCs from unstructured form sources. Several automated methods for extracting IOCs (such as malicious IP addresses, malware, and file hashes of malicious payloads) are based on the OpenIOC standard, including PhishTank, IOCFinder, and CleanMX [14]. To facilitate efficient threat intelligence sharing among organisations, CybOX [22], STIX [23], and TAXII [24] have emerged as de-facto standards for describing threat intelligence and are widely consumed by the threat intelligence sharing platforms, including MISP [12] and AT\&T Open Threat eXchange (OTX).

\subsection{Indicators of Compromise}

Cyber Threat Intelligence (CTI) includes IOCs, which organisations can use to identify possible threats and protect themselves and their customers. Specifically, IOCs are artefacts observed about an attacker or their behaviour, such as tactics, techniques and procedures [25]. IOCs can be kept at a network or host level and help network defenders block malicious traffic and identity actions or determine if a cyber intrusion has occurred. Security and forensic analysts prepare reports of in-depth analysis of cyber attacks, including the IOCs, to be shared with communities, often through public data sources. Examples of IOC found in reports from data sources include actor identity behind cyber attacks, the malware used in threat attacks and their typical behaviour, communication and control server list, and other types of information. The information used in creating these reports is gathered from multiple sources, such as host logs, proxy logs and alerts. The reports may be widely distributed through various channels, including blogs, forums and social media.

The pyramid of pain (PoP) classifies the common types of IOCs. The PoP identifies the types of indicators that a system defender might use to detect an adversary's activities. The pyramid organises the pain an adversary will cause when the defender can deny those indicators. At the bottom end, if a defender identifies hash values of malicious files and then blocks these, it causes the attacker little pain since making an insignificant change to the file to produce the same outcome with a different hash is trivial. TTP sit at the top of the pyramid. When a defender detects and responds at this level, this disrupts behaviours much more complicated for an adversary to change; defining new behaviours is a significant challenge for adversaries.

\subsection{Heterogeneous Information Network}

Heterogeneous Information Network (HIN) is a simple way of modelling a problem as a graph compromising different types of nodes and one or more correlations between nodes (edges) [26]. The set of node and edge types correspond to the network scheme. HIN delivers a high conceptualisation of modelling for a complex collection of data. From the graphical representation of the dataset, feature vectors can be extracted by defining meta-paths and meta-graphs corresponding to the graph and implementing a guided random walk over defined meta-paths and meta-graphs. A meta-path is a path defined within the graph of network schema, covering a specific sequence of relation types. A meta-graph [27] can handle the in-depth relationship between nodes by employing a direct acyclic graph of nodes defined over the HIN from a single source node to a single target node. The guided random walk generates a sequence of nodes processed in an embedding model such as word2vec, skip-gram or Continuous Bag-of-Words (CBOW). Once the nodes are represented numerically, it is possible to determine a set of nodes and resolve many problems (classification, clustering, and similarity search).

\subsection{Overview}

We introduce a novel system designed to automatically extract and prioritise high-level IOCs (Indicators of Compromise) from multiple sources of threat text. Our system addresses the limitations of existing IOC extraction methods by considering the semantic relationships among different IOCs. We present a novel approach to extracting threat-related information and identifying the domains IOCs belong to. This information is then integrated with their respective domains to form three domain-specific threat intelligence categories: the organisational domain, regional-source domain, and regional-target domain. We also present a threat modelling that utilizes a Heterogeneous Information Network (HIN) comprising different meta-paths and meta-graphs. The proposed system captures the interdependent relationships among diverse IOCs and provides a more comprehensive view of the landscape of threat events. Our system then utilizes similarity measures based on these meta-paths and meta-graphs to assess the interdependent relationships among different IOCs.

To prioritize IOCs, we measure their severity by considering various factors, including the frequency of attacks, the lifetime of the IOC, and the number of exploited vulnerabilities in each domain. Our system then evaluates the ranking mechanism for each IOC, providing a more comprehensive and accurate view of the threat landscape. Our system significantly contributes to cybersecurity, providing a more effective and efficient method for automatically extracting, assessing, and prioritizing high-level IOCs. With the increasing frequency and complexity of cyber threats, the need for such a system has become more critical.

\section{Methodology}

The architecture of the proposed method, as shown in Fig 2, comprises of following phases: Data collection and Preprocessing, Relation Extraction and Threat Modelling, Domain Recognition and Tag Generation, Domain-specific threat identification and Tagging, and Severity measure and Threat Prioritisation. Table 1 summarises the list of notations and abbreviations used throughout the paper.

\begin{table}
    \centering
  \caption{List of Abbreviations and Notations}
  \label{tab:tab1}
  \begin{tabular}{cc}
    \hline
    Notations	& Description\\
    \hline
    IOC & Indicators of Compromise \\
    AM & Additive manufacturing \\
    HIN & Heterogeneous Information Network \\
    APT & Advanced Persistent Threats \\
    CTI & Cyber Threat Intelligence \\
    TTP & Tactic, technique and procedure \\
    PoP & Pyramid of pain\\
    STIX & Structured threat information exchange\\
    TAXII & Trusted Automated eXchange of Indicator Information \\
    
    \hline
\end{tabular}
\end{table}

\subsection{Data Collection and Preprocessing}

The system automatically collects threat information identifying IOCs from multiple resources, including forums, blogs, security news, and bulletins. We use a breadth-first search to capture the HTML course code and Xpath for data extraction. We then reduce the dimension of each text report and remove noisy features by pre-processing. This pre-processing includes the removal of stopwords, punctuations, and markup characters. 

\begin{figure} \centering
\includegraphics[width=80mm,scale=0.7]{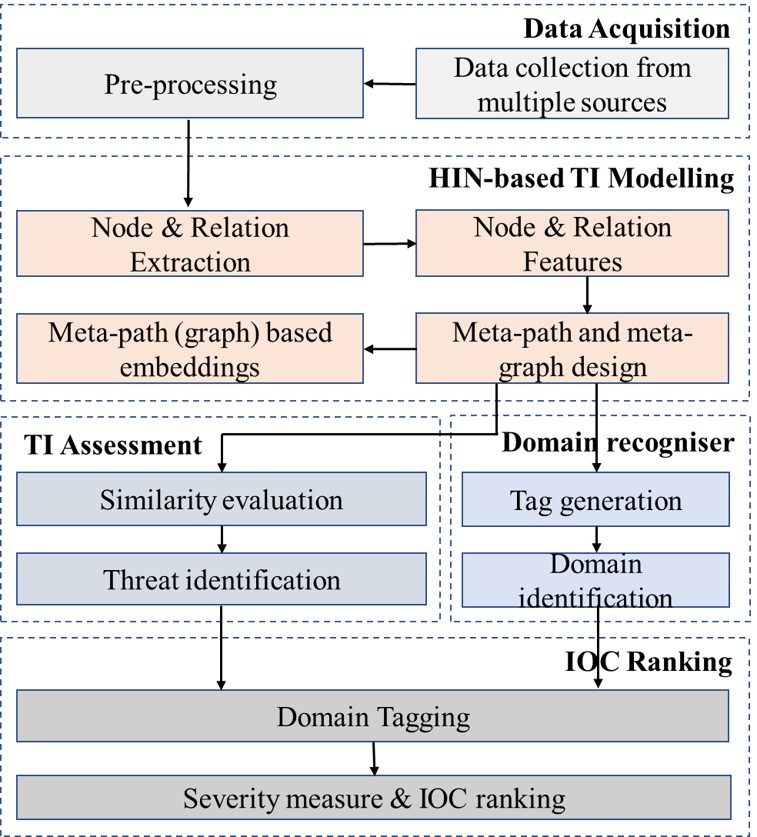}
\caption{Process flow of the Proposed system} \label{fig2}
\end{figure}

\subsection{Relation extraction and threat intelligence modelling}

Using a heterogeneous information network (HIN) for threat intelligence, we first build a graph that shows the interdependent (semantic) relationships between the different IOCs involved in the attack. By denoting nodes (IOCs) and relationships, we can identify patterns and anomalies that may indicate the presence of a threat. For example, we can use HINs to identify groups of attackers that share common attack vectors or targets or to track the evolution of an attack over time as new entities become involved. To better understand, we can characterise the nodes and relations as follows.

\subsubsection{Node Features.}

In the context of risk in Additive Manufacturing, it is essential to consider the domain-specific threat information. For instance, a threat post discussing the Stuxnet virus and its impact on industrial control systems is more relevant to manufacturing organisations than those in the finance or healthcare sectors. This highlights the need for threat intelligence tailored to an organisation's domain.

Additionally, geographical location plays a significant role in cyber attacks. Over 500 geopolitical cyber attacks have been reported worldwide in the past decade, with 30\% originating from China or Russia and 26.3\% targeting the USA. In 2018 alone, 27\% of attacks occurred in the USA [28]. Therefore, when developing threat models for Additive Manufacturing, it is crucial to consider the regional source and target source of cyber attacks.

To account for these domain-specific and regional factors in our threat intelligence model for Additive Manufacturing, we define nodes as organisation-specific, regional\_source-specific, and regional\_target-specific. This enables us to capture the complex relationships between entities involved in cyber attacks, such as attackers, attack vectors, and targets. Moreover, we consider time-related node features such as attack frequency and IOC lifecycle, which can provide valuable insight into the TTPs of attackers and help defenders calculate the level of risk posed by a particular threat.

\subsubsection{Semantic relation features.}

The node features in the HIN represent a specific action, but actions can be employed multiple times in conjunction with other activities in a campaign. These complex relationships among nodes can provide more valuable intelligence for threat identification; therefore, we consider relation-based and node features. This allows us to analyse highly sophisticated malicious cyber attacks. To model the interdependent relationship between eight IOCs, we define the following semantic relationships:

\begin{itemize}
    \item \textbf{R1}: The relation \textbf{actor-exploit-vulnerability} matrix $A$ represents the link between the threat actor and vulnerability. For each element, $A_{i,j}  \in \{0,1\}$, where $A_{i,j}=1$ means actor $i$ exploits vulnerability $j$. 
    
    \vskip 0.3cm
    \item \textbf{R2}: The relation \textbf{actor-invade-device} matrix $B$ represents the link between the threat actor and device. For each element, $B_{i,j}  \in \{0,1\}$, where $B_{i,j}=1$ means actor $i$ invades device $j$. 
    
    \vskip 0.3cm
    \item \textbf{R3}: The link between two actors is represented by the relation \textbf{actor-assist-actor} matrix $C$. For each element, $C_{i,j}  \in \{0,1\}$, where $C_{i,j}=1$ means actor $i$ assists actor $j$. 

    \vskip 0.3cm
    \item \textbf{R4}: The relation \textbf{attack\_type-originate\_from-region} matrix $D$ represents the link between the attack type and location. For each element, $D_{i,j}  \in \{0,1\}$, where $D_{i,j}=1$ means attack type $i$ originate from region $j$. 

    \vskip 0.3cm
    \item  \textbf{R5}: The relation \textbf{attack\_type-target-region} matrix $E$ represents the link between the attack type and location. For each element, $E_{i,j}  \in \{0,1\}$, where $F_{i,j}=1$ means attack type $i$ target to region $j$. 

    \vskip 0.3cm
    \item \textbf{R6}: The relation \textbf{vulnerability-affect-device} matrix $F$ represents the link between the vulnerability and the device. For each element, $F_{i,j}  \in \{0,1\}$, where $F_{i,j}=1$ means vulnerability $i$ affects device $j$.

    \vskip 0.3cm
    \item \textbf{R7}: The relation \textbf{attack\_type-associate-vulnerability} matrix $G$ represents the link between the attack type and vulnerability. For each element, $G_{i,j}  \in \{0,1\}$, where $G_{i,j}=1$ means attack type $i$ carry vulnerability $j$.

    \vskip 0.3cm
    \item \textbf{R8}: The relation \textbf{vulnerability-held-time} matrix $H$ represents the link between the vulnerability and time. For each element, $H_{i,j}  \in \{0,1\}$, where $H_{i,j}=1$ means vulnerability $i$ held in time $j$.

    \vskip 0.3cm
    \item \textbf{R9}: The relation \textbf{vulnerability-include -file} matrix $B$ represents the link between the vulnerability and malicious file. For each element, $I_{i,j}  \in \{0,1\}$, where $I_{i,j}=1$ means vulnerability $i$ include malicious file $j$.

    \vskip 0.3cm
    \textbf{R10}: The relation \textbf{vulnerability-evolve-vulnerability} matrix $B$ represents the link between the vulnerabilities. For each element, $J_{i,j}  \in \{0,1\}$, where $J_{i,j}=1$ means vulnerability $i$ evolve to vulnerability $j$.
\end{itemize}

We initiate dependency parsing to leverage the semantic relationships among the eight IOCs and extract them in a structured format. Using this approach, we can represent the IOCs as triplets, each consisting of two IOCs and a relation between them. For instance, if IOC1 is dependent on IOC2, we would define the relationship as (IOC1-relation-IOC2), where 'relation' denotes the nature of the relationship between the two IOCs.

\subsubsection{Meta-path and Meta-graph.}

Figure 3 presents 12 distinct types of meta-paths and meta-graphs denoted by $\chi_i$ that capture interdependent relationships among seven different IOCs. While the meta-path illustrates the connections between the IOCs, it falls short in capturing intricate relationships. To address this limitation, the proposed HIN-based Threat Intelligence (TI) model utilizes a directed acyclic graph of nodes to handle more complex structures in the HIN architecture. By learning and analyzing these 12 different meta-paths and meta-graphs, the model can convey the context of a threat event and offer threat insights across heterogeneous IOCs. For instance, the $\chi_1$ meta-path is a length-2 meta-path that represents the relatedness of "threat actors (A) exploiting the same vulnerability (V)."

\begin{figure} \centering
\includegraphics[width=100mm,scale=0.7]{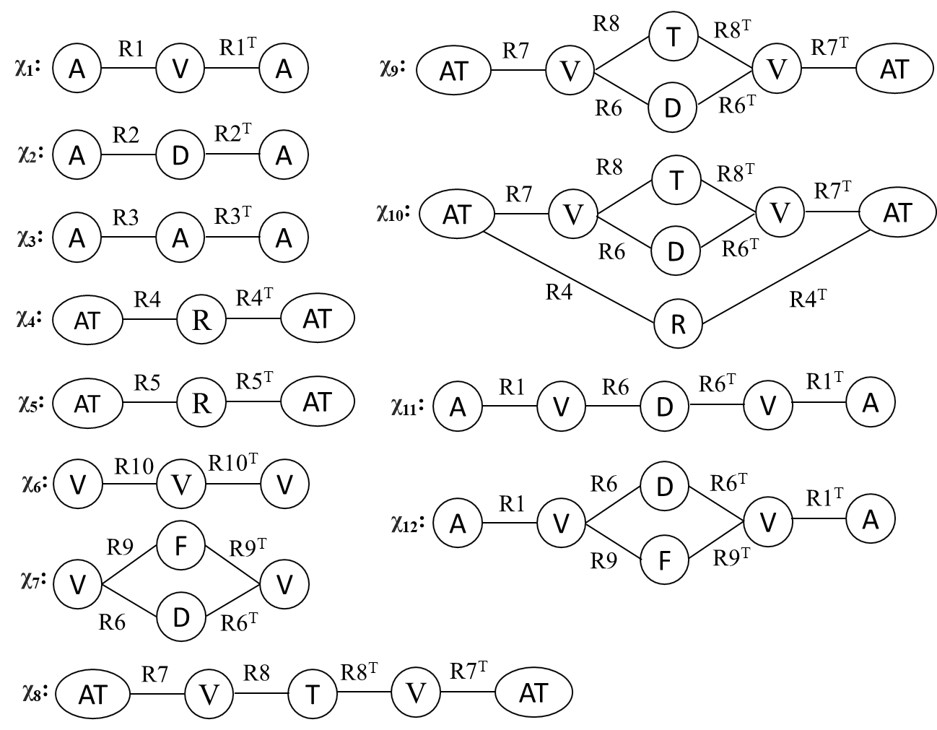}
\caption{Proposed meta-path and meta-graph for threat type identification, where A denotes threat actor, V denotes vulnerability, D denotes device, R denotes a region, AT denotes attack type, F denotes file, and T denotes time.} \label{fig3}
\end{figure}

Similarly, $\chi_8$ is a meta-path that describes the relationships between IOCs that “two attack types who leverage the same vulnerability held at the same time”. Likewise, $\chi_{10}$ is a meta-graph that portrays the relationship over threat infrastructure with more comprehensive insight that integrates both external and intrinsic relationships. Meta-graph $\chi_{10}$ depicts the relationship among IOCs: "two attack types originated from the same region, and their associated vulnerabilities affect the same device occur at the same time”.

\subsection{Domain recognition and tag generation}

To extract domain-specific IOCs, it is essential first to identify the domain of threat information. This initial step helps to ensure that IOCs are tailored to the specific context of the threat landscape, enabling more effective threat detection and response. Here, we consider three domains, \textit{organisation\_domain-specific}, \textit{regional\_source-specific}, and \textit{regional\_target-specific}. \textit{Organisational\_domain-specific} threat information includes financial, health, manufacturing, government, and IoT information. \textit{Regional\_source-specific} and \textit{Regional\_target-specific} threat information originated and targeted the geographic region, such as China, Russia, India, Korea, USA, UK, and Europe. 
We first trained the word2vec model specific to a threat description embedding that inputs a large corpus (threat description) and generates a low-dimension vector. Each unique word in the corpus is allocated an assigned vector in latent space. The convolution function sets a filter for each word vector to generate a feature called local\_feature. Our model captures the most significant features by leveraging the max-pooling operation that takes the maximum values of local-feature and runs over a local feature set. This will generate tags for three domains, i.e., $OD_t$, $SD_t$, and $TD_t$ denotes the tags corresponding to organisation\_domain-specific, regional\_source-specific, and regional\_target-specific threat, respectively.

\subsection{Domain-Specific Threat Identification and Tagging}

After successfully extracting the features of IOCs and their relationships and identifying the relevant meta-paths and meta-graphs, a meta-path and meta-graph-based heterogeneous classifier is developed to classify the threat type of infrastructure nodes in Cyber Threat Intelligence (CTI). The proposed classification approach integrates node features and explores the semantic similarities between meta-paths and meta-graphs to represent the nodes comprehensively. These advanced techniques enable a more comprehensive depiction of the nodes, enhancing the accuracy of the threat classification.

Given the threat intelligence graph $G=(N,R)$, and meta-path and meta-graph set $P=\{\chi_1,\chi_2,..\chi_n\}$, the assessment of threat intelligence includes the following steps:

\begin{itemize}
    \item \textbf{Adjacency matrix}.     The relationships between threat nodes can be explored using different meta-paths and meta-graphs, which capture the behaviour of threats in various aspects. To represent these relationships, we propose using an adjacent weighted matrix, denoted by $Adj_i \in R^{N \times N}$, which can be generated using similarity algorithms such as Euclidean distance, Manhattan distance, Cosine similarity, word mover distance, and Jaccard similarity. To assess the similarity between IOCs, we generate a corresponding weighted adjacency matrix, $Adj_i$, based on the meta-path and meta-graph path set, $P$. The use of weighted adjacency matrices $Adj_i$ enables the identification of the most significant relationships between the different nodes, which can be used to prioritise threat mitigation efforts.

    \item \textbf{Feature matrix}.    By incorporating attributed information of nodes, we can construct an attributed feature matrix $F_i$ of size $N \times d$, where N denotes the number of IOCs in $Adj_i$, and $d$ is the dimension of the node feature. This allows us to integrate the attribute information of IOCs and create a node feature matrix $F_i \in R^{N \times d}$. To recognize previously unnoticed IOCs, we employ the word2vec method to develop a threat intelligence embedding, which transforms words into a latent vector space. To achieve this, threat-related texts are pre-processed, accumulated into a word set, and converted into a latent vector space using word2vec. This approach enables us to represent threat-related information in a low-dimensional vector space, facilitating the detection and analysis of potential threats.
    
    \item \textbf{Quantify threat intelligence}. After designing an adjacent weighted and attributed feature matrix, we assess the threat intelligence. Different types of assessment methods to quantify the proposed HIN-based threat intelligence. For example, graph convolution network (GCN) and Bidirectional Encoder Representations from Transformers (BERT). Given the adjacency matrix $Adj_i$, and its corresponding feature matrix $F_i$ (low-dimensional space), we utilise the graph convolution network (GCN) method to quantify the relationship between IOCs. This will fuse the adjacency matrix $Adj_i$, and feature matrix $F_i$ as $Z=(F,Adj)$ and output the predicted labels of IOCs. Then, the model integrates the domain-specific tags $OD_t$, $SD_t$ and $TD_t$ to the predicted IOC, representing that the IOCi belongs to the organisation $OD_t$, originates from the country $SD_t$ and is targeted to the country $TD_t$ and are considered as the domain-specific IOCs. 
\end{itemize}

\subsection{Severity Measure and Threat Prioritisation} Utilizing the learned domain-specific IOCs, we can evaluate the severity of potential threats of various attack vectors within each domain. This motivates us to develop a quantitative measure to assess threat risks corresponding to each domain. The proposed severity measure is based on several key assumptions.

\begin{enumerate}
    \item Firstly, we assume that the frequency of attacks may significantly influence the severity and scope of the threats manifested. 
    \item Secondly, we postulate that chain exploits, where multiple vulnerabilities use an attack, can cause considerably more damage. 
    \item Finally, we recognize that the severity of a threat may decrease over time, particularly during the zero-day risk period.
\end{enumerate}

Consequently, the severity of a threat can be measured by examining the frequency of attacks, the lifetime of the IOC, and the number of exploited vulnerabilities in each domain. This approach allows us to develop a more nuanced and comprehensive understanding of the potential threats facing a domain, enabling us to take appropriate measures to mitigate risk and enhance security.

\section{Conclusion and Future scope}
This paper presented a novel semantic-based threat prioritisation system for AM that intends to expose the comprehensive behaviour of threat events in the relationships among different IOCs with high accuracy. We proposed an intelligent IOC acquisition and ranking system based on a Heterogeneous Information Network (HIN). The proposed system collects threat-related from multiple sources that automatically extract threat-related information, measure the severity of IOCs, and quantify them based on the severity. We considered individual IOCs and one or more relationships among semantically similar IOCs. We proposed an efficient recogniser to identify domain-specific IOCs focusing on three domains: \textit{organisational\_domain-specific, regional\_source-specific}, and \textit{regional\_target-specific} threat intelligence. Further, we evaluated the severity of IOC by exploring the frequency of attacks, IOC lifetime, and the number of exploited vulnerabilities in each domain.  

The proposed semantic-based threat prioritisation system for AM has potential future scopes that can be explored, such as:

\begin{itemize}
    \item \textit{Integrating with existing security tools}: The proposed system can be combined with existing security tools to provide real-time threat intelligence and prioritisation of threats. The integration can help security teams to automate the detection, investigation, and response to threats and reduce the time to mitigate them.
    \item \textit{Exploring additional domains}: The proposed system focuses on three domains: \textit{organisational\_domain-specific}, \textit{regional\_source-specific}, and \textit{regional\_target-specific}. However, other domains, such as \textit{industry-specific} or \textit{technology-specific}, can be explored to provide a more comprehensive view of the threat landscape.
    \item \textit{Improving the ranking system}: The proposed system ranks the IOCs based on severity. However, the ranking system can be improved to consider the evolving threat landscape and real-time threat intelligence data to enhance the accuracy of the prioritisation system.
\end{itemize}

\end{document}